\documentclass[prl,twocolumn,amsmath,amssymb]{revtex4}
\usepackage{graphicx}

\usepackage{dcolumn,amsfonts}
\begin{document}

\title{Competing ferromagnetism in high temperature copper oxide superconductors}
\author{Angela Kopp}
\author{Amit Ghosal}
\author{Sudip Chakravarty}

\affiliation{Department of Physics and Astronomy, University of California, Los Angeles, California 90095-1547, USA}

\date{\today}

\begin{abstract}
\bf{The extreme variability of observables across the phase diagram of the cuprate high temperature superconductors has remained a profound mystery, with no convincing explanation of the superconducting dome. While much attention has been paid to the underdoped regime of the hole-doped cuprates because of its proximity to a complex Mott insulating phase, little attention has been paid to the overdoped regime. Experiments are beginning to reveal that the phenomenology of the overdoped regime is just as puzzling. For example, the electrons appear to form a Landau Fermi liquid, but this interpretation is problematic; any trace of Mott phenomena, as signified by incommensurate antiferromagnetic fluctuations, is absent, and the uniform spin susceptibility shows a ferromagnetic upturn. Here we show and justify that many of these puzzles can be resolved if we assume that competing ferromagnetic fluctuations are simultaneously present with superconductivity, and the termination of the superconducting dome in the overdoped regime marks a quantum critical point beyond which there should be a genuine ferromagnetic phase at zero temperature. We propose new experiments, and make new predictions, to test our theory and suggest that effort must be mounted to elucidate the nature of the overdoped regime, if the problem of high temperature superconductivity is to be solved. Our approach places competing order as the root of the complexity of the cuprate phase diagram.} 
\end{abstract}
\maketitle

\begin{figure}[htb]
\begin{center}
\includegraphics[width=0.4\textwidth]{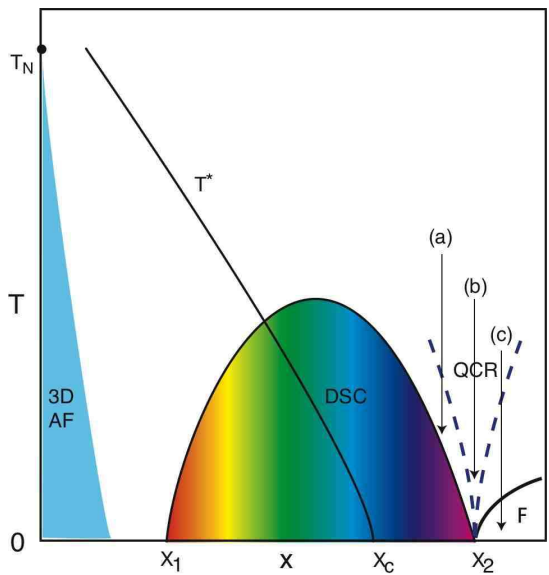}
\caption{A sketch of a  simplified  phase diagram of the cuprates as a function of temperature, $T$, and carrier concentration (doped holes), $x$. The superconducting dome rises at about $x_{1}\approx 5$\% and ends at about $x_{2}\approx 30$\%; $x_{1}$ and $x_{2}$ are two proposed QCPs terminating the $d$-wave superconductor (DSC). The regime proximate to $x_{1}$ is termed underdoped and that close to $x_{2}$ is designated overdoped.  The enigmatic  pseudogap state is marked by the line $T^{*}$. A plausible candidate for this state at low temperatures  is the $d$-density wave (DDW)~\cite{Chakravarty:2001}. The doping  $x_{c}$ marks another QCP, and roughly the region between $x_{c}$ and $x_{2}$ is the overdoped regime. The underdoped regime is riddled with other non-superconducting competing phases, such as stripes, checkerboard charge order, spin glass, and the three-dimensional antifrromagnet (3D AF). Here $x_{2}$ marks the formation of a ferromagnetic (F) phase. The quantum critical regime emanating from $x_{2}$ is labelled QCR. The arrows (a), (b), and (c) are possible experimental trajectories. It is only for (c) that a static ferromagnetic order parameter can develop}\label{Fig1}
\end{center}
\end{figure}

The superconducting dome (See Fig.~\ref{Fig1}.), that is, the shape of the superconducting transition temperature $\mathrm{T_{c}}$ as a function of doping (added charge carriers), $x$, is a clue that the high-$\mathrm{T_{c}}$ superconductors are  unconventional.
Conventional superconductors, explained so beautifully by Bardeen, Cooper, and Schrieffer (BCS) have a unique ground state that is not naturally separated by any non-superconducting states. The electron-phonon mechanism leads to superconductivity for arbitrarily weak attraction between electrons. To destroy a superconducting state requires a magnetic field or strong material disorder. In the absence of disorder or magnetic field, it is difficult to explain the sharp cutoffs at $x_{1}$ and $x_{2}$ within the BCS theory. For high-$\mathrm{T_{c}}$ superconductors, there is considerable evidence that  competing  order parameters  are the underlying reason. Thus, $x_{1}$ and $x_{2}$ signify quantum phase transitions, most likely quantum critical points (QCP).~\cite{Sachdev:1999} Understanding  high-T$_{c}$ superconductors therefore require an understanding of possible competing orders.~\cite{Sachdev:2003,Chakravarty:2001} The QCP at $x_{1}$ is extensively studied,~\cite{Balents:1998,Herbut:2005,Kopp:2005,Franz:2006,Barzykin:2006} but little is known about $x_{2}$. Recent work has also emphasized the importance of the maximum of the uniform susceptibility in defining the pseudogap line $T^{*}$ in Fig.~\ref{Fig1} that ends at another QCP at $x_{c}$.~\cite{Barzykin:2006} Here, we attempt, instead, at gaining insight from the possible existence of a QCP at $x_{2}$.

Complex materials---cuprates, heavy fermions, organics--- are rife with spontaneously broken symmetries in their ground states. This is unsurprising from a theoretical perspective. Because two distinct macroscopic states cannot be analytic continuations of each other,  they cannot be literally said to arise from each other. Macroscopic properties are best described by a low-energy effective Hamiltonian. In the process of coarse-graining, one generates effective Hamiltonians with a multitude of coupling constants that signify the intermediate processes removed from the original Hamiltonian.  Tuning these coupling constants can easily lead to a variety of ordered states that compete with each other. Thus, it is a truism that any symmetry that can be broken, must be broken. Competing phases can therefore explain the rich  phase diagram of the high temperature superconductors.

Nonetheless, it is frequently assumed that the superconductivity in the overdoped cuprates (See Fig.~\ref{Fig1}.) grows out of a Landau Fermi liquid. Experimental support has included the observation of angular magnetoresistance oscillations in $\mathrm{Tl_{2}Ba_{2}CuO_{6+\delta}}$ (Tl2201)~\cite{Hussey:2003} and the reported verification of the Wiedemann-Franz law (WF) in both Tl2201 and $\mathrm{La_{2-x}Sr_{x}CuO_{4}}$ (LSCO),~\cite{Proust:2002,Nakamae:2003} materials that can be overdoped to the edge of the superconducting dome. The Fermi liquid interpretation has not been without difficulties, however.~\cite{Abdel-Jawad:2006} The verification of WF in Ref.~\cite{Proust:2002} can be questioned, as the relation was found to coexist with a large linear resistivity term extending down to zero temperature. Given a number of difficulties, this is not clear  even for Ref.~\cite{Nakamae:2003}.   An example is the failure of the Kadowaki-Woods (KW) relation.~\cite{Kadowaki:1986} The resistivity, $\rho$ , of a Fermi liquid, no matter how strongly interacting it may be, is given by $\rho=\rho_{0}+AT^{2}$, $T\to 0$, where $\rho_{0}$   is its residual value and $T$  is the temperature; similarly, its specific heat is given by $C=\gamma_{0}T$, $T\to 0$. The relation between $A$  and $\gamma_{0}$ is known as the KW relation, and it is easy to see that the KW ratio, $A/\gamma_{0}^{2}$ , should be independent of the effective mass $m^{*}$, that is, the band mass $m$ renormalized by interactions. The striking enhancement~\cite{Nakamae:2003} of the KW ratio in heavily overdoped LSCO and Tl2201 over the ``universal'' heavy fermion $(m^{*}\gg m)$ value, but with $m^{*}\approx m$ , is surprising. The anomalous character of the overdoped $\mathrm{La_{1.7}Sr_{0.3}CuO_{4}}$ is also clear~\cite{Nakamae:2003} from the extremely narrow range of temperature for which the linearity of specific heat holds, while the quadratic temperature dependence of the resistivity holds over a surprisingly broad temperature range up to ~50 K, beyond which it behaves as $T^{1.6}$. There is little doubt that electron-electron interaction cannot be subsumed by Fermi liquid renormalizations, and the measured enhancement of the KW ratio is not particularly meaningful. Clearly, further experiments are necessary.

\begin{figure}[htb]
\begin{center}
\includegraphics[width=0.4\textwidth]{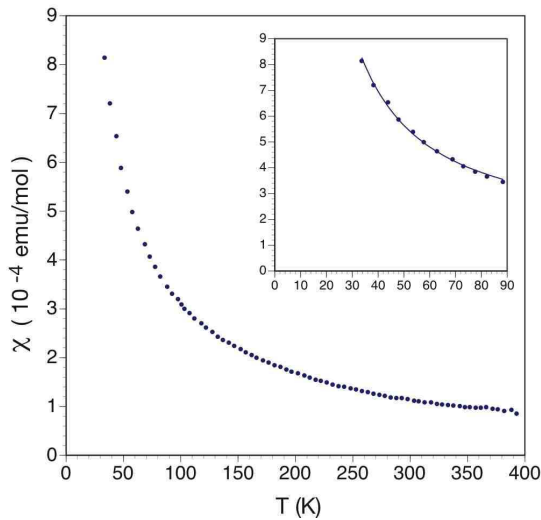}
\caption{Total magnetic susceptibility in a field of 1T of Tl2201 in the overdoped regime with $T_{c}$ tuned to zero (data excerpted from Ref.~\cite{Kubo:1991}). The data can be fit to the form $\chi=A+B/T^{\gamma}$ globally over the entire range. The core diamagnetic susceptibility, $-1.93\times 10^{-4}\textrm{emu/mol}$, is independent of temperature.  The  range between 30 K and 100 K can be equally well fit (within the accuracy of the data) with  $\gamma$ between $0.95-1.25$. The inset shows a fit with $\gamma=5/4$, a prediction of the quantum critical scaling given in the text. Clearly quantum critical scaling should become more precise at asymptotically low temperatures for which data are not available. See Ref.~\cite{Kubo:1991} for other hole concentrations. Similar data are also available for LSCO and LBCO (See Refs.~\cite{Oda:1990,Wakimoto:2005}).}\label{Fig2}
\end{center}
\end{figure}

	The centerpiece of the anomalies is the uniform magnetic susceptibility, which does not remotely resemble that of a Fermi liquid. It is known that in $\mathrm{La_{2-x}M_{x}CuO_{4}}$  (M=Sr, Ba), the susceptibility exhibits a dramatic upturn at temperatures just above the transition temperature $T_{c}$   for $x>0.2$, while for lower doping it behaves as it would if the ground state were a singlet. Similarly, overdoped Tl2201 shows a striking upturn in susceptibility; see Fig.~\ref{Fig2}. In both cases, the upturn becomes more pronounced with overdoping.~\cite{Kubo:1991,Oda:1990,Wakimoto:2005,Takagi:1989,oda:1991,Kondo:1994} Despite motivated effort, it has not been possible to attribute this unusual behaviour to Curie susceptibility of a magnetic impurity phase.  An explanation in terms of localized moments, as suggested by Nakano {\em et al.}~\cite{Nakano:1994}, is also problematic, because the effective Curie constant would have to increase with hole doping, when the system becomes more metallic. This seems very unlikely. Moreover, a careful analysis shows that $1/4$ of $\mathrm{Sr}$ ions in $\mathrm{La_{2-x}Sr_{x}CuO_{4}}$ corresponding to holes that exceed $x=0.22$ have to create paramagnetic moments, which is equally unlikely.~\cite{Wakimoto:2005} Neutron scattering measurements around $q=(\pi,\pi)$ in the same sample show anomalous enhancement of dynamic antiferromagnetic correlations due to $\mathrm{Zn}$ substitution, while   
$\mathrm{Sr}$ overdoping results in a smooth decrease of the imaginary part of the dynamic susceptibility. Given that   $\mathrm{Zn}$ substitution is known produce paramagnetic moments, the different behavior of  $\mathrm{Sr}$ overdoping argues against the formation of paramagnetic moments.

One of the hallmarks of special correlation effects that give rise to Mott insulators (as opposed to band insulators) is antiferromagnetism, which takes on an incommensurate character with doping. Recent neutron scattering experiments in LSCO indicate, however, that the intensity of the finite energy incommensurate antiferromagnetic peaks diminishes with increased doping, vanishing as the material is tuned to the edge of the superconducting dome at  $x=0.3$.~\cite{Wakimoto:2004} This is a striking testament to the disappearance of these correlations in the overdoped regime.

Here we suggest that ferromagnetic fluctuations compete with $d$-wave superconductivity in the overdoped regime. Ferromagnetism can be a consequence of strong electronic correlations and low particle density (large hole density) when  Fermi surface nesting is remote or the system is not close to a Mott insulating phase.~\cite{Fazekas:1999,Honerkamp:2001} Anderson~\cite{Anderson:2002} has also noted the possible role of ferromagnetic correlations in the overdoped regime from the perspective of Nagaoka mechanism~\cite{Becca:2001,Putikka:1992} approaching it from the underdoped side. We consider, instead, the itinerant picture and use it not to justify the pseudogap crossover, but the rich phenomenology of the overdoped cuprates. It would be interesting to study the intersection of the two ideas. We suppose that $x_{2}$  (See Fig.~\ref{Fig1}.) signifies either a QCP or a first order quantum phase transition between a $d$-wave superconductor and a ferromagnetic ground state. A first order transition can become continuous in the presence of disorder---for order parameters of continuous symmetry, this is expected to occur in dimensions $d\le 4$,~\cite{Imry:1979,Hui:1989,Aizenman:1989,Aizenman:1990,Berker:1993} which can be straightforwardly extended to a quantum phase transition between two ordered states. Thus, ferromagnetic fluctuations in the superconducting state will act as pair breakers and explain the demise of superconductivity in overdoped cuprates. Of course, pair breaking will occur even if the transition is weakly first order---large ferromagnetic domains will still appear in the superconducting phase. 

The possibility of the coexistence of two phases would have been plausible if the superconductivity were of the triplet variety. In principle ferromagnetic fluctuations can give rise to spin-triplet $p$-wave superconductivity, but this appears to be far too delicate.~\cite{Monthoux:2004} In any case, at least approximately up to 
22\% doping, phase sensitive experiments seem to indicate robust $d$-wave pairing in the cuprates~\cite{Tsuei:2004}, which is also generally believed to be true within the entire superconducting dome.

It was argued elsewhere that the criticality at $x_{1}$   is that of a quantum three-dimensional  $XY$-model,~\cite{Sachdev:1999} the fundamental variable being the phase of the superconducting order parameter. This alone leads to the robust result $T_{c}\propto \rho_{s}(0)^{1/2}$, where $\rho_{s}(0)$ is the  $T=0$  superfluid density.~\cite{Kopp:2005} Recent experiments~\cite{Broun:2005} are consistent with this quantum critical result asymptotically close to $x_{1}$, showing, a strong deviation from the widely assumed Uemura plot, $T_{c}\propto  \rho_{s}(0)$.~\cite{Uemura:1989}

What would be the corresponding result at  $x_{2}$, if it is a QCP as well? For $|x-x_{2}|\ll 1$, pair breaking will lead to $T_{c}\propto  |x-x_{2}|^{\theta}$, where $\theta$  is an exponent to be determined. Because of spin diffusion, the dynamic exponent, $z$, for a dirty metallic ferromagnet is given by 
$z=4$,~\cite{Fulde:1968} dissipation being relevant on both sides of the transition; on the superconducting side it arises from the gaplessness. From an interesting experiment~\cite{Cox:1988} on a classical system, it appears that a disorder driven continuous transition can be of emergent $\phi^{4}$ -type (Ising),~\cite{Zinn:2002} while the actual phases on either sides of the transition are determined by the operators that are irrelevant at the critical point.  We surmise that the same is true for the quantum problem, except for an added dissipative term in the effective action.~\cite{Hertz:1976,Millis:1993} Since for $d=3$, $(d+z)>4$, $\rho_{s}(0)\sim |x-x_{2}|$, the mean field result. Due to the dangerous irrelevant operator, $u$  , controlling the interaction of the bosonic modes, a one-loop calculation (higher loops cannot change the leading scaling result) leads to the correlation length $\xi(T, x=x_{2}) \sim 1/T^{(z+1)/2z}$, consistent with Ref.~\cite{Millis:1993}. But as the phase boundary is given by $\xi(T=T_{c},x=x_{2})\sim \xi(T=0,x)$, the exponent $\theta$  is given by $\theta=z/(z+1)$, using the mean field result $\xi(T=0,x)\sim |x-x_{2}|^{-1/2}$. If we substitute $z=4$, we get $\theta=4/5$  and $T_{c}\propto \rho_{s}(0)^{4/5}$, in contrast to the underdoped regime, providing an explanation of the Boomerang effect of  $T_{c}$ vs. $\rho_{s}(0)$  discussed earlier,~\cite{Niedermeyer:1993} where pair breaking of unspecified origin was suggested. This exponent can be tested in future experiments.  (The exponent in the overdoped regime derived in Ref.~\cite{Kopp:2005} would be 2, which is incorrect.) 

Recently an empirical  scaling analysis~\cite{Homes:2004} shows that $\rho_{s}(0)\propto \sigma_{\textrm{dc}}(T\approx T_{c})T_{c}$ holds reasonably well for a broad range of doping, both for in-plane and out-of-plane components, where the quantity $\sigma_{\textrm{dc}}(T\approx T_{c})$ is the conductivity at $T\approx T_{c}$. A similar scaling relation was derived earlier for the out-of-plane components.~\cite{Chakravarty:1999} Even if this relation were to hold for the entire doping range, this would not invalidate the Boomerang effect, which is the statement that the $T=0$ exponents of  $\rho_{s}(0)$, as a function of $x$, are different at the QCPs $x_{1}$ and $x_{2}$. This would merely mean that these exponents are the sum of the exponents of the conductivity and the transition temperature as a function of $x$ at the respective QCPs.

Ferromagnetic fluctuations result in the surprising upturn of the susceptibility. Previous fits to the data have assumed a  $1/T$-dependence.  For a first order transition, this may be an adequate description of the contribution from finite-size ferromagnetic domains.  But if a QCP exists at $x_{2}$, the susceptibility, $\chi$, will exhibit quantum critical scaling; for $x$  tuned to $x_{2}$, $\chi\sim \xi(T,x=x_{2})^{2}\sim T^{-\gamma}$,  as $T\to 0$, where $\gamma = 5/4$ ($z=4$).  This is consistent with the data shown in Fig.~\ref{Fig2} for Tl 2201, which, to a good approximation, is tuned to $x_{2}$. Note that $\chi$   approaches the Curie form as  $z\to \infty$ and $\theta\to 1$.  Importantly, for the range of doping explored in experiments thus far, the data cannot, {\em even in principle}, reflect a static ferromagnetic order; for this, the doping must be larger than  $x_{2}$ (See Fig.~\ref{Fig1}.). 
From a one-loop theory (similar to the susceptibility calculation) we predict that for $x=x_{2}$  the magnetization, $M$, and the applied field, $h$, must obey (also approximately true in the entire quantum critical region):
\begin{equation}
M=(h/u)^{1/3}f(uh^{2}\chi^{3})
\label{eq:M}
\end{equation}
where the universal scaling function $f(y) \to 1$ as $y\to \infty$, and $f(y)\to y^{1/3}$ as $y\to 0$. Here $d=3$ and $(d+z)>4$. The crossover from the finite temperature result to the zero temperature result is given by $\chi h \sim (h/u)^{1/3}$. Since presently there are no known samples for $x>x_{2}$, we can only verify  this quantum critical scaling by either susceptibility measurements down to low temperatures, or by {\em precise field dependence of Knight shift}, a measurement that to our knowledge has not been carried out.  

It is useful to sketch this calculation in a bit more detail. Consider a $N$-component $\phi^{4}$ theory with dissipation corresponding $z=4$ such that $(d+z)>4$, that is, above its upper critical dimension.  We integrate out the Matsubara frequencies $\omega_{n}\ne 0$ to construct an effective $d$-dimensional theory.~\cite{Sachdev:1999,Zinn:2002} The leading scaling result can be obtained from a one-loop calculation; higher order calculations can not change the leading behavior. The spatially independent part of the quadratic term, $R$, and the coefficient of the quartic term, $U$, are
\begin{eqnarray}
R&=&r+u\left(\frac{N+2}{6}\right)T\sum_{n\ne 0}\int \frac{d{\bf q}}{(2\pi)^{d}}\chi({\bf q},i\omega_{n}),\\
U&=& u+{\cal O}(u^{2}),
\end{eqnarray}
where $\chi^{-1}({\bf q},i\omega_{n})=q^{2}+|\omega_{n}|/Dq^{2}+r$, and $r\propto (x_{2}-x)$ is the original quadratic term in the quantum $(d+z)$-dimensional problem. Here $D$ is the spin diffusion constant of an itinerant metallic ferromagnet. There is of course no renormalization of the term containing the static magnetic field $h$. Tuned to the $T=0$ quantum critical point at $r=0$, the equation for the magnetization becomes,
\begin{equation}
\xi^{-2}(r=0,T)M - h +\frac{U}{3!}M^{3}=0,
\label{eq:cubic}
\end{equation}
where 
\begin{equation}
\xi^{-2}(r=0,T)=u\left(\frac{N+2}{6}\right)T\sum_{n}\int \frac{d{\bf q}}{(2\pi)^{d}}\chi_{0}({\bf q},i\omega_{n}),
\end{equation}
and $\chi^{-1}_{0}({\bf q},i\omega_{n})=q^{2}+|\omega_{n}|/Dq^{2}$. Note that the sum now includes $n=0$. The solution of the cubic equation in Eq.~\ref{eq:cubic} now gives the universal scaling function in Eq.~\ref{eq:M}.
\begin{figure}[htb]
\begin{center}
\includegraphics[width=0.4\textwidth]{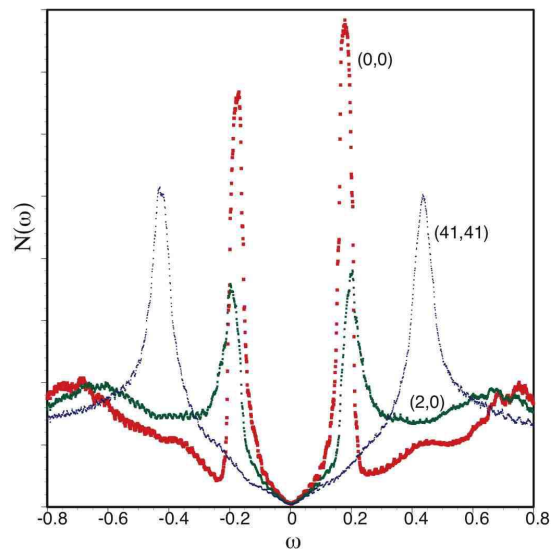}
\caption{The local density of states (LDOS)   (arb. units) as a function of frequency $\omega$, $N({\bf r},\omega)$, in a $d$-wave superconductor (DSC) with a central ferromagnetic domain of size $5\times 5$   embedded in a sample of size  $41\times 41$ (in terms of lattice units). The curves labelled $(m,n)$ correspond to the location where the density of states is calculated from Bogoliubov-de Gennes theory. The density of states at $(0,0)$  consists of an extraordinarily sharp peak with a small gap. As we move out of the centre, these peaks decrease in intensity, get wider, and eventually the LDOS corresponding to the surrounding DSC take over. This effect is demonstrated by calculating LDOS at $(2.0)$  and $(41,41)$. We see sharp peaks in the central region  even though the depression of the order parameter due to ferromagnetism is only 5\%. Note also the change in the low energy spectra as we move out of the central region, which becomes linear, as it should for a pure DSC. The parameters used in this calculation are listed in the methods section.}\label{Fig3}
\end{center}
\end{figure}

As to excitations in the superconducting state, quantum critical fluctuations will primarily scatter the nodal quasiparticles as $x\to x_{2}$, because their energy approaches zero; in degenerate perturbation theory, modes of equal energies couple most strongly.  Thus, we expect nodal quasiparticles to broaden with overdoping while the antinodal quasiparticles, whose energy is finite, will sharpen, because they will not couple strongly to the critical fluctuations. This surprising effect may have already been observed in recent angle resolved photoemission measurements.~\cite{Plate:2005} This is in part due to the remarkable effort in recent years in the preparation of high quality single crystals of overdoped Tl2201.~\cite{Peets:2006} At present two models involving impurity scattering may be relevant,~\cite{Zhu:2004,Wakabayashi:2005} although these predict a isotropic quasiparticle  scattering rate above $T_{c}$ and a temperature dependence set by the scale of the energy gap. More experiments are necessary to discriminate between different ideas, but such substantial impurity scattering in the high quality single crystals seems unlikely. As to inelastic mechanism, another possibility is the proximity of a competing $d_{x^{2}-y^{2}}+id_{xy}$ superconducting phase, similar to our suggestion of a competing ferromagnetic state.~\cite{Vojta:2000}

The STM (scanning tunnelling microscopy) spectra should be sensitive to the inhomogeneity due to ferromagnetic domains. We used a self-consistent Bogoliubov-de Gennes theory to calculate the local density of states (LDOS) for a $d$-wave superconductor modified to include an effective Zeeman field. The calculational method has been described in detail in the past.~\cite{Ghosal:2005} Here we simply note the modifications necessary for the present problem. The Hamiltonian should contain an added ferromagnetic term $H_{F}= -J_{F}\sum_{\langle ij\rangle}{\bf S}_{i}\cdot {\bf S}_{j}$, where the sum is over nearest neighbour pairs of spin operators ${\bf S}_{i}$. We approximate this term in mean field theory as a Zeeman field  $-h\sum_{i} S_{i}^{z}$. While $h$  is treated as a fixed quantity, all other parameters are determined self-consistently. The  remaining Hamiltonian is a $t-t'-J$-model, where the units are such that $t=1$, $t'=-0.27$. $J=1.27$. These are effective parameters including mean field renormalizations (Gutzwiller-constraint) due to hole doping of $\delta=25$\%. These parameters result in a uniform DSC with a $d$-wave gap $\Delta_{k}= (\Delta_{0}/2)(\cos k_{x} - \cos k_{y})$, where $\Delta_{0} =0.44$. For Fig.~\ref{Fig3}, we chose $h/\Delta_{0} =0.4$; the reduction of the self-consistent pairing amplitude at the centre of the ferromagnetic patch with respect to  $\Delta_{0}$ was about 5{\%}. The size of the unit cell was  $41\times 41$ replicated in the repeated zone scheme $14\times 14$   times. The central ferromagnetic patch was of size $5\times 5$. We have repeated the calculation with a variety of parameters. The results shown in Fig.~\ref{Fig3} are representative. The ferromagnetic exchange constant,  $J_{F}$, corresponding to the assumed Zeeman field is much smaller than the ferromagnetic $J_{0}$  that one needs to explain the phenomena in the overdoped regime in terms of magnetic impurity effects.
The field, which is zero everywhere except for a small patch of sites at the centre of the lattice, represents a pair-breaking ferromagnetic domain. As shown in Fig.~\ref{Fig3}, the results of this calculation are consistent with the recent work where the gap in the central patch was arbitrarily set to zero.~\cite{Fang:2006} We see the same effect even though the amplitude is reduced by only 5\%, provided self-consistency is imposed.

We emphasize, however, that the gap inhomogeneity in the underdoped regime~\cite{McElroy:2005} is entirely different.~\cite{Ghosal:2005} There the large gap regions reflect a pseudogap due to a competing orde---in our opinion, the $d$-density wave~\cite{Chakravarty:2001}---while the narrow gap regions reflect the $d$-wave superconductor. This change in character of the elementary excitations should take place around the middle of the dome (See Fig.~\ref{Fig1}.).

If ferromagnetism is truly a competing phase in the overdoped regime, the vortices would be easier to nucleate and the physics of the vortex core could be dramatically different, which could be probed in STM and other measurements. The most exciting test of our proposal will be a direct observation of the ferromagnetic phase, which would require the fabrication of samples with hole concentrations greater than $x_{2}$.  Since there can be no ferromagnetic order at finite temperatures in two dimensions, the Curie temperature will be set by the interlayer coupling and is likely to be very small, especially for a disordered ferromagnet with direct exchange or dipolar coupling.  Nonetheless, we hope that our ideas are sufficiently tantalizing to spark the interest of the experimentalists.

\begin{acknowledgments}
We acknowledge support from the U. S. National Science Foundation and from funds from the David Saxon Chair at UCLA. We thank N. Peter Armitage, Stuart Brown, Steven Kivelson, Alessandara Lanzara,   A. Peter Young,  and especially Elihu Abrahams for comments. 

\end{acknowledgments}

\end{document}